\documentclass[aps,preprint,superscriptaddress,nofootinbib]{revtex4}
\usepackage[utf8]{inputenc}
\usepackage{amsmath,amssymb,bbm}
\usepackage{dsfont}
\usepackage{graphicx}
\usepackage{xcolor,slashed}
\pdfoutput=1
\usepackage{appendix}

 \usepackage{hyperref}       
 \usepackage{url}            
 \usepackage{booktabs}       
 \usepackage{amsfonts}       
 \usepackage{nicefrac}       
 \usepackage{microtype}      
 \usepackage{lipsum}

\begin{document}
\rightline{CQUeST-2026-0802}
\bigskip
\bigskip

\title{Probing Charged Higgs Bosons through $H^\pm \to W^\pm A$ Decay \\ in the Type-I Two-Higgs Doublet Model}

\author{Abdesslam Arhrib}
\email[E-mail: ]{aarhrib@uae.ac.ma}
\affiliation{Abdelmalek Essaadi University, FST Tanger B.P. 416, Morocco}
\author{Mohamed Krab}
\email[E-mail: ]
{mkrab@hep1.phys.ntu.edu.tw}
\affiliation{Department of Physics, National Taiwan University, Taipei 106319, Taiwan}
\author{Tanmoy Mondal}
\email[E-mail: ]
{tanmoy.mondal@pilani.bits-pilani.ac.in}
\affiliation{Birla Institute of Technology and Science, Pilani, 333031, Rajasthan, India}
\author{Prasenjit Sanyal}
\email[E-mail: ]
{prasenjit.sanyal01@gmail.com}
\affiliation{Center for Quantum Spacetime, Sogang University, Seoul 121-742, South Korea}
\affiliation{Department of Physics, Sogang University, Seoul 121-742, South Korea}

\begin{abstract}
A light pseudoscalar lighter than the 125~GeV Higgs boson remains allowed within the type-I Two-Higgs Doublet Model (2HDM), consistent with current collider constraints, including Higgs signal strength measurements and direct searches for additional Higgs bosons.  
At the Large Hadron Collider (LHC), the light pseudoscalar $A$ can be pair-produced through the {\it undetected} decay of the Standard Model-like Higgs boson, $h_{125} \to AA$, when this channel is kinematically accessible (i.e., $m_A < m_{h_{125}}/2$).
For the charged scalar~$H^\pm$, once open, the bosonic decay channel $H^\pm \to W^{\pm(*)} A$ naturally dominates over the fermionic $H^\pm \to \tau^\pm\nu$ and $H^\pm \to tb$ channels, while relaxing the existing $H^\pm$ limits.
As a result, we propose searching for the charged Higgs boson through $H^\pm \to W^{\pm(*)} A$ in the $4b+2j+\ell\nu$ final state.
This channel is particularly sensitive to moderate-to-large $\tan\beta$ values and provides an excellent complementary probe to searches based on fermionic $H^\pm$ decays at the LHC.
Our collider analysis demonstrates that this channel can probe a substantial region of the type-I 2HDM parameter space at the high-luminosity LHC.
\end{abstract}

\maketitle

\label{intro}
\section{Introduction}		

The discovery of a Higgs boson with a mass of 125 GeV by the ATLAS~\cite{ATLAS:2012yve} and CMS~\cite{CMS:2012qbp} collaborations at the Large Hadron Collider (LHC) marked a milestone in particle physics and completed the particle spectrum predicted by the Standard Model (SM). Subsequent measurements of its production cross sections and decay rates have demonstrated remarkable agreement with SM expectations, within the current experimental uncertainties \cite{ATLAS:2017ovn,ATLAS:2019nkf}. 
Nevertheless, establishing the nature of  Electroweak Symmetry Breaking (EWSB) mechanism and probing possible deviations from the SM requires precision studies of Higgs couplings to gauge bosons and fermions.
 Despite the success of the SM, several theoretical and phenomenological puzzles remain unresolved, including the hierarchy problem, the origin of neutrino masses, the nature of dark matter, and the observed matter–antimatter asymmetry of the Universe~\cite{Langacker:2008yv,Bertone:2004pz,KamLAND:2002uet,Bonilla:2019ipe,Ma:2006km}. 
These unresolved  issues provide compelling motivation for exploring extensions of the SM, particularly those featuring an enlarged scalar sector that could accommodate new phenomena  and distinctive experimental signatures.

Among the simplest and most extensively studied extensions of the SM Higgs sector is the Two-Higgs Doublet Model (2HDM) \cite{Lee:1973iz,Glashow:1976nt,Branco:2011iw}. By introducing an additional $SU(2)_L$ scalar doublet, the 2HDM enlarges the Higgs spectrum and predicts five physical scalar states after EWSB: 
two CP-even neutral Higgs bosons ($h$ and $H$), one CP-odd pseudoscalar ($A$), and a pair of charged Higgs bosons ($H^\pm$). 
The discovery of a charged Higgs boson would constitute unambiguous evidence of physics beyond the SM.
Consequently, searches for charged Higgs bosons represent an important component of the physics programmes at the LHC and future colliders.

The production and decay properties of charged Higgs boson depend sensitively on their masses and couplings, which, in turn, are determined by the specific realisation of the 2HDM. Extensive direct searches have been carried out 
at LEP~\cite{ALEPH:2013htx,OPAL:2008tdn}, the Tevatron~\cite{CDF:2005acr,D0:2009vff}, and the LHC~\cite{ATLAS:2012nhc,ATLAS:2012tny,ATLAS:2016avi, ATLAS:2014otc, ATLAS:2018gfm,CMS:2012fgz,CMS:2014cdp,CMS:2015lsf,CMS:2019bfg,CMS:2018ect,ATLAS:2024hya,ATLAS:2013uxj,CMS:2015yvc,CMS:2020osd,ATLAS:2024oqu,CMS:2018dzl,ATLAS:2023bzb,CMS:2014pea,ATLAS:2015nkq,ATLAS:2018ntn,CMS:2019rlz,CMS:2020imj,ATLAS:2021upq,CMS:2025plw} 
placing significant constraints  on the charged Higgs boson mass. 
At LEP, pair production via $e^+e^- \to H^+ H^-$ led to lower limits on the charged Higgs mass around 80 GeV, assuming dominant decays into fermionic channels  \cite{ALEPH:2013htx,OPAL:2008tdn}. At hadron colliders, when the charged Higgs mass satisfies $m_{H\pm}<m_t+m_b$, the dominant production mechanism typically 
proceeds via top-quark pair production, followed by the decay $t\to b H^+$. 
This channel has been extensively studied at the LHC, leading to stringent limits on the product: $\mathrm{BR}(t\to bH^+) \times \mathrm{BR}(H^+ \to \tau^+ \nu)$~\cite{ATLAS:2024hya}. 
For heavier charged Higgs states, direct production processes such as $bg\to tH^+$ and $gg\to tbH^+$ become relevant \cite{Barger:1993th,Diaz-Cruz:1992ust,Borzumati:1999th}.
In this work, we consider three mass regimes for charged Higgs production: light, intermediate, and heavy, as discussed in detail in Section~\ref{sect:Hproduction}.
In addition, several alternative production mechanisms have been investigated in the literature. 
These include associated production with a $W$ boson, $pp\to W^\pm H^\mp$ \cite{ Dicus:1989vf,BarrientosBendezu:2000tu,Brein:2000cv,Bao:2010sz}, 
production in association with a neutral Higgs boson $pp\to H^\pm S$~($S = h, H, A$) \cite{Kanemura:2001hz,Cao:2003tr,Akeroyd:2003bt,Akeroyd:2003jp,Belyaev:2006rf,Miao:2010rg,Enberg:2018pye, Arhrib:2017wmo,Arhrib:2021xmc,Wang:2021pxc,Mondal:2021bxa,Kim:2023lxc,Mondal:2023wib,Sanyal:2023pfs}, and charged Higgs pair production via quark annihilation or gluon fusion,  
$pp\to H^\pm H^\mp$ \cite{BarrientosBendezu:1999gp,Krause:1997rc}. Although these processes generally exhibit smaller cross sections compared to top-associated production in certain regions of parameter space, they can become significant under some specific threshold effects or coupling configurations~\cite{Arhrib:2021xmc}.

The decay pattern of a charged Higgs boson is equally important in determining its experimental signatures and the strategies used to search for it. In the 2HDM, fermionic decay modes such as $H^\pm \to \tau^\pm \nu$, $H^\pm \to cs$, and, when kinematically allowed,  
$H^\pm \to tb$  have traditionally been 
considered dominant, especially in type-II 2HDM scenarios (see, e.g., Ref.~\cite{Akeroyd:2016ymd}). 
However, in extended scalar sectors where additional neutral Higgs states are lighter than the charged Higgs, bosonic decay channels such as 
$H^\pm\to W^\pm h, W^\pm H$ and/or $H^\pm \to W^\pm A$ can become competitive or even dominant in substantial regions of parameter space \cite{Arhrib:2016wpw,Bahl:2021str}. 
Such decays open complementary discovery channels and can weaken the applicability of existing experimental constraints obtained under the assumption of predominantly fermionic charged Higgs decays. 
Despite their potential importance, experimental searches for bosonic charged Higgs decays $H^\pm\to W^\pm h, W^\pm H, W^\pm A$, remain comparatively limited~\cite{CMS:2019idx,ATLAS:2021xhq,CMS:2022jqc,ATLAS:2024rcu}. This motivates a dedicated investigation of these channels and their discovery prospects at the LHC.

In the present work, we focus on the $H^\pm \to W^{\pm} A$ decay channel and its subsequent contribution to the $4b+2j+\ell\nu$ final state. 
We consider the type-I 2HDM, where a light charged Higgs boson can remain compatible with flavor physics constraints and existing collider data, we first identify the viable parameter space by imposing theoretical and experimental constraints. We then investigate the collider phenomenology of this channel, evaluating its discovery potential at the LHC through a dedicated signal-to-background analysis. Our results demonstrate that 
$H^\pm \to W^{\pm} A$ channel offers a promising complementary probe to searches based on fermionic charged Higgs decays at the LHC.

The remainder of this paper is organized as follows. Section~\ref{sect:model} introduces the type-I 2HDM and summarizes its relevant features. Section~\ref{sect:scan} presents the parameter space analysis, incorporating  theoretical and experimental constraints. Section~\ref{sect:Hproduction} discusses charged Higgs production and decay at the LHC. Section~\ref{sect:analysis} presents our collider analysis of the $H^\pm \to W^{\pm} A$ channel. Finally, our conclusions are summarized in Section~\ref{sect:concl}.
		
\section{The type-I 2HDM} 
\label{sect:model}

The most general CP-conserving 2HDM potential is given by \cite{Branco:2011iw}:
\begin{eqnarray}
V &=& m_{11}^2(\Phi_1^\dagger\Phi_1) +
m_{22}^2(\Phi_2^\dagger\Phi_2) -
[ m_{12}^2(\Phi_1^\dagger\Phi_2)+\text{H.c.}] ~\nonumber\\&& 
+ \frac12\lambda_1(\Phi_1^\dagger\Phi_1)^2 +
\frac12\lambda_2(\Phi_2^\dagger\Phi_2)^2 +
\lambda_3(\Phi_1^\dagger\Phi_1)(\Phi_2^\dagger\Phi_2) ~\nonumber\\ && +
\lambda_4(\Phi_1^\dagger\Phi_2)(\Phi_2^\dagger\Phi_1) +
\frac12\left[\lambda_5(\Phi_1^\dagger\Phi_2)^2 +\rm{H.c.}\right],
\label{CTHDMpot}
\end{eqnarray}
where all parameters are taken to be real. Using the two minimization conditions of the scalar potential, $m_{11}^2$ and $m_{22}^2$ can be traded for the vacuum expectation values $v_{1,2}$ of the Higgs doublets $\Phi_{1,2}$. The quartic couplings $\lambda_{1-5}$ can then be expressed in terms of the four physical Higgs masses, the mixing angles $\alpha$ and $\beta$, and the parameter $m_{12}^2$. The scalar sector can thus be parametrized by seven independent parameters: $m_h$, $m_H$, $m_A$, $m_{H^\pm}$, $\sin(\beta-\alpha)$, $\tan\beta$, and $m_{12}^2$.

To avoid tree-level flavor-changing neutral currents (FCNCs) in the Yukawa sector, a discrete $Z_2$ symmetry can be imposed, ensuring that fermions couple to only one of the two Higgs doublets~\cite{Glashow:1976nt}. We allow the $Z_2$ symmetry to be softly broken by the $m_{12}^2$ term in the scalar potential.

Depending on the Yukawa coupling assignments, four types of 2HDM can be distinguished \cite{Branco:2011iw}. In the present study, we only consider Yukawa texture of type-I where $\Phi_2$ doublet couples to all the SM fermions exactly like in the SM.


In the type-I 2HDM, the couplings of the neutral and charged Higgs bosons to fermions are given by \cite{Branco:2011iw} 
\begin{eqnarray}
{\mathcal{L}}_{\rm Yukawa} &=& - \sum_{f=u,d,l} \frac{m_f}{v} \left(\frac{c_\alpha}{s_\beta} \bar{f} f h + \frac{s_\alpha}{s_\beta} \bar{f} f H 
- i\,\mathrm{sgn}(Q_f) \cot\beta \, \bar{f} \gamma_5 f A \right) \\
&& - \left\{ \frac{\sqrt{2} V_{ud}}{v} \bar{u} \left( m_u \cot\beta P_L -
m_d \cot\beta P_R \right) d H^+ - \frac{\sqrt{2} m_\ell}{v} \cot\beta\,\bar{\nu}_L \ell_R H^+ + \text{H.c.} \right\}, \nonumber
\label{Yukawa-1}
\end{eqnarray}
where $c_x \equiv \cos x$, $s_x \equiv \sin x$, $\mathrm{sgn}(Q_f) = +1~(-1)$ for $f=u$~($f = d, \ell$), $V_{ud}$ denotes CKM matrix element, and $P_{L,R} = (1 \mp \gamma_5)/2$. The coupling of the two CP-even states $h$ and $H$ to gauge bosons $W,Z$ are controlled by $s_{\beta-\alpha}$ and $c_{\beta-\alpha}$, respectively. In the alignment limit, the couplings of the CP-even Higgs bosons to gauge bosons approach their SM values when $c_{\beta-\alpha} \rightarrow 0$ (for a SM-like $h$) or $s_{\beta-\alpha} \rightarrow 0$ (for a SM-like $H$),
corresponding to normal and inverted hierarchies, respectively.
Consequently, both scenarios can accommodate the 125~GeV Higgs signal observed at the LHC. 
In this study, we identify the observed 125~GeV state with the lighter CP-even scalar, $h$ (normal hierarchy).

The couplings of the charged Higgs boson to the $W$ boson and the neutral Higgs bosons $h$, $H$, and $A$ are given by
\begin{eqnarray}
    C(H^\pm W^\mp h) &:& \mp i\frac{g}{2} c_{\beta-\alpha} ( p^\mu_h - p^\mu_{H^\pm} ), \\
    C(H^\pm W^\mp H) &:& \pm i\frac{g}{2} s_{\beta-\alpha} ( p^\mu_H - p^\mu_{H^\pm} ), \\
    C(H^\pm W^\mp A) &:& \frac{g}{2}  ( p^\mu_A - p^\mu_{H^\pm} ),   
\end{eqnarray}
where $g$ is the $SU(2)_L$ gauge coupling. In the alignment limit of $c_{\beta-\alpha} \to 0$, $C(H^\pm W^\mp h)$ is suppressed, while $C(H^\pm W^\mp H)$ is maximized. The coupling $C(H^\pm W^\mp A)$ is independent of the mixing angle $\beta-\alpha$ and thus not affected by the alignment limit.

\section{Scan of the type-I 2HDM parameter space}
\label{sect:scan}
In this study, we fix $m_h = 125$~GeV and randomly scan the remaining model parameters in the ranges given in Table~\ref{table:scan}.
The scan is performed using the code \texttt{ScannerS}~\cite{Muhlleitner:2020wwk}, where all parameter points are required to satisfy the following theoretical and experimental constraints:

\begin{itemize}
   \item Perturbative unitarity, boundedness from below~\cite{Branco:2011iw}, and vacuum stability~\cite{Barroso:2013awa}.
	
   \item Electroweak precision constraints through the oblique parameters $S$, $T$, and $U$~\cite{Peskin:1990zt,Peskin:1991sw,Grimus:2007if,Grimus:2008nb}, using the following fit results obtained under the assumption $U=0$~\cite{Fischer:2026bka}:\footnote{The 2HDM contribution to $U$ is generally suppressed and is therefore neglected.} 
   \begin{equation}
       S = 0.01 \pm 0.07,\quad T = 0.02 \pm 0.05,\quad \rho_{ST} = 0.94,
   \end{equation}
   where $\rho_{ST}$ is the correlation factor between $S$ and $T$.

   \item Flavor constraints using the fit results of Ref.~\cite{Haller:2018nnx}.

   \item Exclusion bounds from additional Higgs boson searches at collider experiments using the code \texttt{HiggsBounds}~\cite{Bechtle:2020pkv} via \texttt{HiggsTools}~\cite{Bahl:2022igd}.
   
   \item Agreement with 125 GeV Higgs signal measurements using the code \texttt{HiggsSignals}~\cite{Bechtle:2020uwn} via \texttt{HiggsTools}~\cite{Bahl:2022igd}. 

\end{itemize}

Note that the perturbative unitarity, boundedness-from-below, vacuum-stability, and flavor constraints are checked using \texttt{ScannerS}. 
The oblique parameters $S$ and $T$, together with the Higgs branching ratios, are also computed with \texttt{ScannerS} and subsequently passed to our code to check the electroweak precision and Higgs constraints.
%
\begin{table}[t]
	\centering
	\setlength{\tabcolsep}{10pt}
	\begin{tabular}{c c c c c c c} \hline
            & $m_H~[\rm{GeV}]$ & $m_A~[\rm{GeV}]$ & $m_{H^\pm}~[\rm{GeV}]$ & $c_{\beta-\alpha}$ & $\tan\beta$ & $m^2_{12}~[\rm{GeV}^2]$ \\ \hline
           min & 130 & 30  & 130 & $-0.3$ & 1  & 0  \\ \hline
           max & 300 & 120 & 300 & 0.3 & 50 & $10^5$ \\ \hline		
	\end{tabular}
	\caption{Type-I 2HDM input parameter ranges.} 
	\label{table:scan}
\end{table} 

The presence of a light CP-odd Higgs boson $A$, with $m_A<m_h/2$, opens up the exotic decay channel $h\to AA$.
This decay is subject to stringent constraints from direct searches (see Ref.~\cite{CMS:2024uru}) and precision measurements of the properties of the observed 125~GeV Higgs boson~\cite{ATLAS:2022vkf,CMS:2022dwd}. 
The corresponding partial decay width at tree level is
given by
\begin{equation}
\Gamma(h\to AA)
=
\frac{g_{hAA}^2}{32\pi m_h}
\sqrt{1-\frac{4m_A^2}{m_h^2}},
\label{eq:Gamma_hAA}
\end{equation}
where $g_{hAA}$ denotes the triple scalar coupling.
In the CP-conserving 2HDM, this coupling can be expressed as~\cite{Bahl:2021str}
\begin{equation}
g_{hAA}
=
-\frac{1}{v}
\left[
(m_h^2+2m_A^2-2M^2)s_{\beta-\alpha}
+
2(m_h^2-M^2)c_{\beta-\alpha}\cot(2\beta)
\right],
\label{eq:ghAA}
\end{equation}
where $M^2=m_{12}^2/(s_\beta c_\beta)$.
Since our analysis focuses on the $H^\pm\to W^\pm A$
decay channel, we tune  $m_{12}^2$ to suppress the $hAA$ coupling, ensuring that the $h\to AA$ decay branching ratio remains compatible with experimental constraints.

\begin{figure}[t]
	\centering
	\includegraphics[scale=0.5]{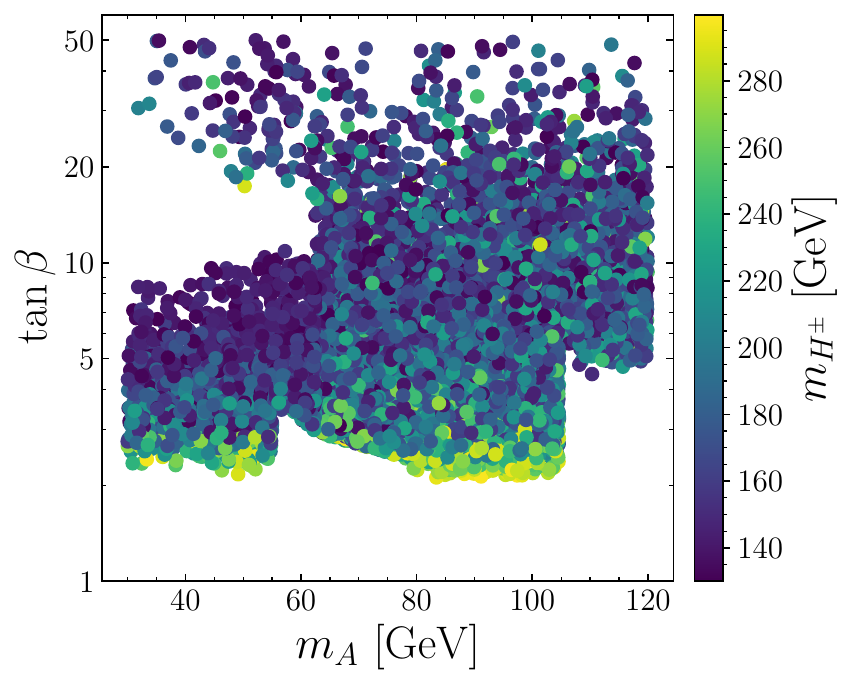}
	\includegraphics[scale=0.5]{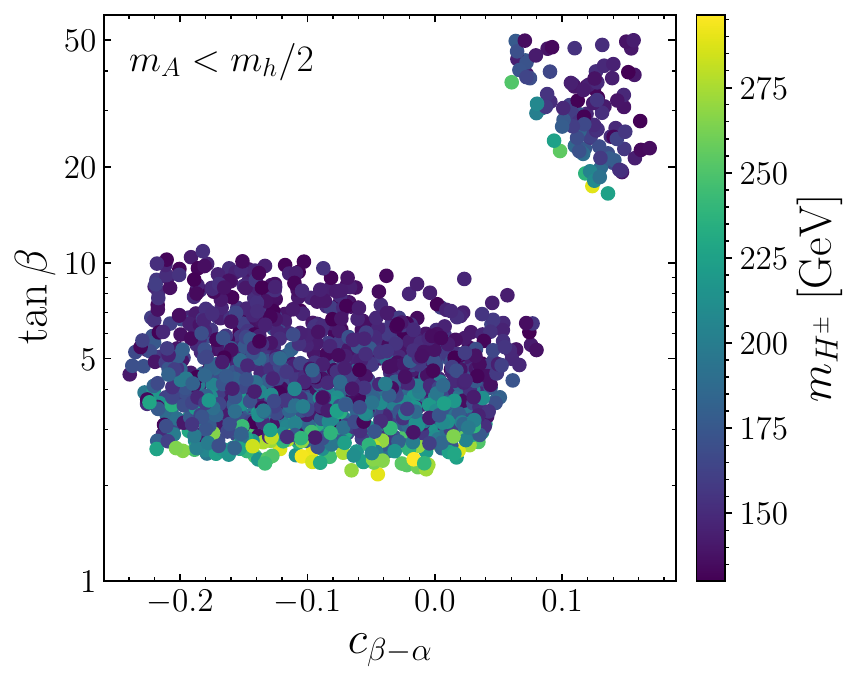}   
	\includegraphics[scale=0.5]{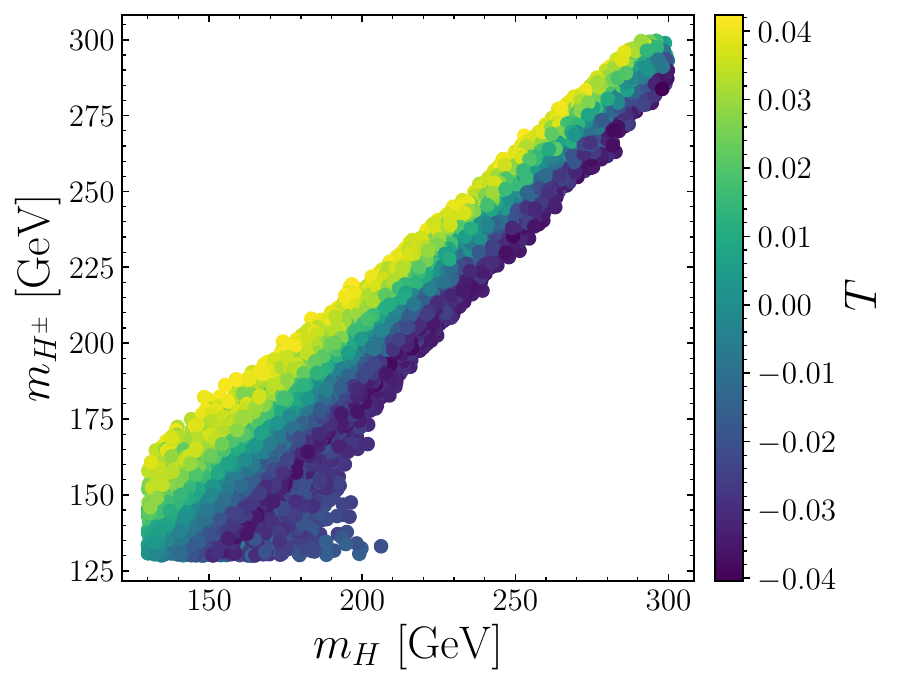}
	\caption{Allowed parameter points in the $m_A$--$\tan\beta$ (upper left), $c_{\beta-\alpha}$--$\tan\beta$ (upper right), and $m_H$--$m_{H^\pm}$ (bottom) planes. The upper-right panel is restricted to $m_A<m_h/2$. The charged Higgs boson mass $m_{H^\pm}$ is indicated by the color bar in the upper panels, while $T$ parameter is shown in the bottom.}
	\label{fig:mA-tb}
\end{figure}

The parameter points satisfying all the constraints discussed above are shown in Fig.~\ref{fig:mA-tb}. For $m_A<m_h/2$, the suppression of the $hAA$ coupling allows viable parameter points to be found at large $\tan\beta$. In this region, the allowed parameter space in the $c_{\beta-\alpha}$--$\tan\beta$ plane (right panel) exhibits two disconnected branches. Perturbative unitarity strongly constrains the large $\tan\beta$ region for negative $c_{\beta-\alpha}$, while measurements of the 125~GeV Higgs boson, as implemented in \texttt{HiggsSignals}, further restrict the parameter space at intermediate $\tan\beta$. Nevertheless, a viable region at large $\tan\beta$ survives, predominantly for positive $c_{\beta-\alpha}$.

\section{Charged Higgs Production and Decay}
\label{sect:Hproduction}

\subsection{Production}

The QCD production of a charged Higgs boson at the LHC can be categorized into light, intermediate, and heavy mass regimes, based on the relative importance of resonant top-quark production and associated production with heavy quarks~\cite{Sanyal:2019xcp}. For a light charged Higgs boson ($m_{H^\pm}\lesssim150$~GeV), the dominant contribution arises from $pp\to t\bar t$, followed by $t\to bH^+$. The on-shell top decay is kinematically open for $m_{H^\pm}<m_t-m_b$, and sufficiently below this threshold, the narrow-width approximation allows one to factorize the charged Higgs production cross section into the top-pair production rate and the corresponding top-decay branching fraction. The $pp\to t\bar t$ cross section is computed at NNLO in QCD using \texttt{Top++2.0}~\cite{Czakon:2011xx}.

For the intermediate regime,
$150~\mathrm{GeV}\lesssim m_{H^\pm}\lesssim200~\mathrm{GeV}$,
the phase space for $t\to bH^+$ becomes very small, and we consider the process $pp\to H^\pm W^\mp b\bar b$, including double-resonant, single-resonant, and non-resonant top-quark contributions as well as their interference~\cite{Degrande:2016hyf,Sanyal:2019xcp}. This approach incorporates finite-top-width effects and connects the contribution from on-shell top decay to that from associated production. The production cross section in this intermediate region is estimated in the four-flavor scheme (4FS), where the bottom quark is treated as massive and does not contribute to the proton parton distribution functions.

For the heavy charged Higgs regime ($m_{H^\pm}>200$~GeV), top-associated production becomes the dominant production mode. The production cross section in this regime is estimated in the 4FS and five-flavor scheme (5FS) schemes~\cite{Flechl:2014wfa,Degrande:2015vpa,LHCHiggsCrossSectionWorkingGroup:2016ypw}, 
and combined to obtain the total cross section using the Santander matching scheme~\cite{Harlander:2011aa} for different values of $\tan\beta$. 


\begin{figure}[t]
\centering
\includegraphics[width=8cm]{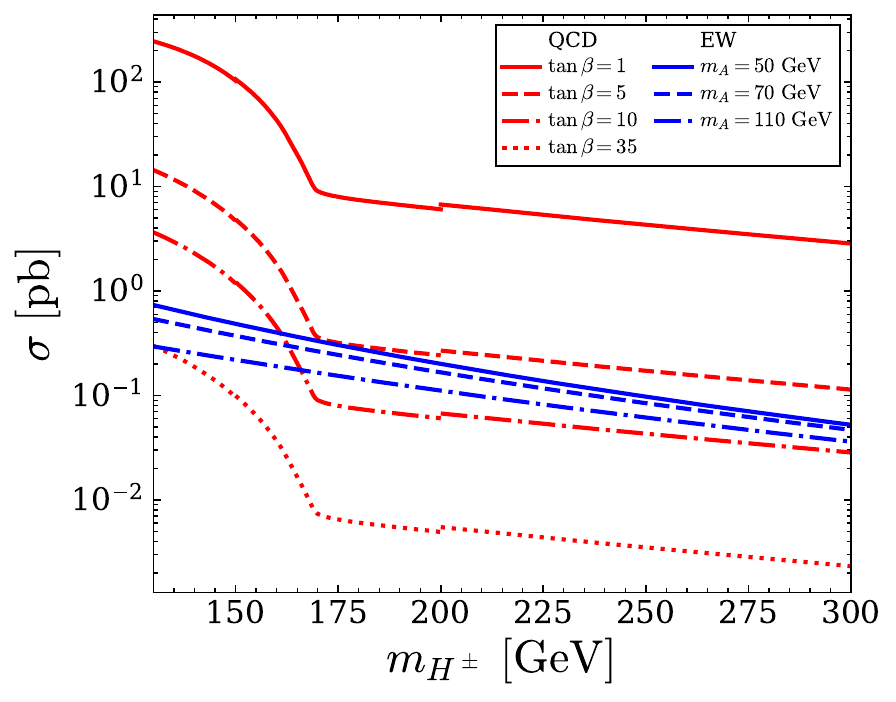}
\caption{Charged Higgs production cross sections at the 13 TeV LHC as
functions of $m_{H^\pm}$. The red curves show QCD-induced production
for $\tan\beta=1$ (solid), $5$ (dashed), $10$ (dash-dotted) and $35$ (dotted).
The blue curves show electroweak (EW) associated production
$pp\to H^\pm A$ for $m_A=50$~GeV (solid), $70$~GeV (dashed),
and $110$~GeV (dash-dotted).
}
\label{fig:Hp-production}
\end{figure}

The QCD-induced production rates  at 13 TeV LHC are shown by the red curves in Fig.~\ref{fig:Hp-production},  where solid, dashed, 
dot-dashed and dotted curves correspond to $\tan\beta = 1, ~5, ~10$ and 35 respectively. The cross section falls sharply as $m_{H^\pm}$ approaches the top-decay threshold, reflecting the disappearance of the resonantly enhanced $t\bar t$ contribution. Above this region, the decrease is much slower and follows the mass dependence of the associated production process. As discussed above, we use the 4FS up to $m_{H^\pm}=200$~GeV and switch to the combination of 4FS and 5FS for heavier charged Higgs masses. The discrete jump at $200$~GeV originates from this change of schemes and does not correspond to a physical threshold.

In the presence of a light pseudoscalar, an additional electroweak production channel,
$q\bar q'\to W^{\pm *}\to H^\pm A$, becomes viable. This process is governed by the gauge coupling and is independent of both $\tan\beta$ and $\beta-\alpha$. In contrast, the charged Higgs Yukawa couplings in the type-I 2HDM are proportional to $\cot\beta$, so that the QCD-induced production rate decreases approximately as $1/\tan^2\beta$.\footnote{In the intermediate region, this simple scaling is modified when $\mathrm{BR}(t\to bH^+)$ becomes sizable, since the charged Higgs decay contributes to the total top width. For our analysis we follow Ref.~\cite{Degrande:2016hyf} for the scaling of cross-section in this regime.} 

The corresponding electroweak production rates are shown by the blue curves in Fig.~\ref{fig:Hp-production} where the solid, dashed and dot-dashed blue curves correspond to pseudoscalar masses of 50, 70 and 110 GeV, respectively. At low $\tan\beta$, QCD-induced production remains the dominant source of charged Higgs bosons, whereas, for large $\tan\beta$, the electroweak $H^\pm A$ channel can become dominant. The two production mechanisms are therefore complementary and become relevant in different regions of the parameter space. Fig.~\ref{fig:Hp-production} shows that for a charged Higgs heavier than the top quark, electroweak production dominates over the QCD production for $\tan\beta$ between 5 and 10. For a charged Higgs lighter than the top quark, the QCD production cross section increases significantly, such that electroweak production becomes dominant only at very large values of $\tan\beta \gtrsim 35$.

\subsection{Decay}
\begin{figure}[t]
\centering
\includegraphics[width=\textwidth]{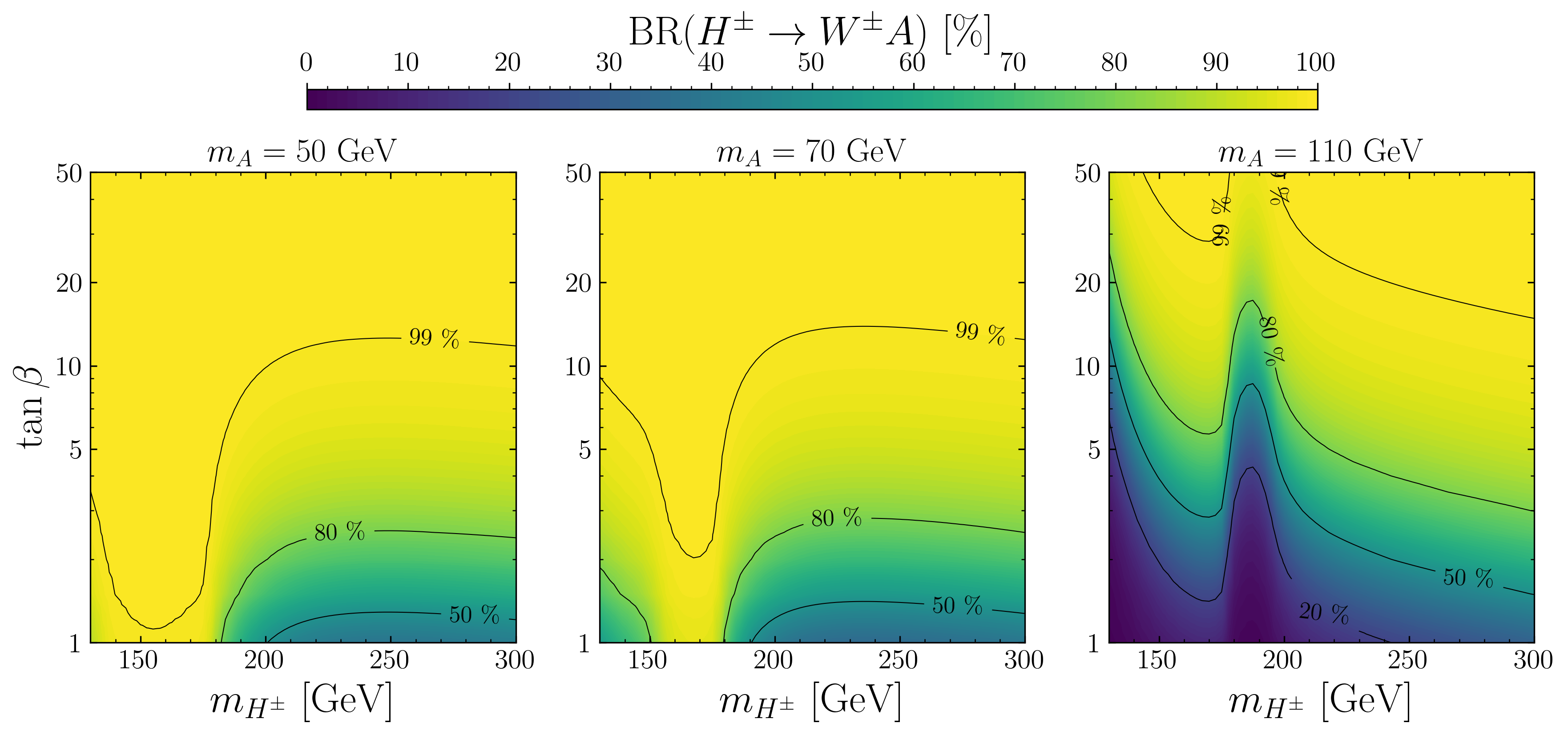}
\caption{Branching fraction of $H^\pm\to W^{\pm(*)}A$ in the $m_{H^\pm}$--$\tan\beta$ plane for $m_A=50$~GeV (left),
$70$~GeV (middle), and $110$~GeV (right), with
$m_H=m_{H^\pm}$, $m_h=125$~GeV, and
$\sin(\beta-\alpha)=1$. The color scale and labelled contours
give the branching fraction in percent.}
\label{fig:BRHpWA}
\end{figure}

The presence of a light pseudoscalar can significantly modify the decay pattern of the charged Higgs through the bosonic mode $H^\pm\to W^\pm A$. In our analysis, we consider three representative values of the pseudoscalar mass, $m_A=50$, $70$, and $110$~GeV. We also set $m_H=m_{H^\pm}$ and impose the alignment limit $\sin(\beta-\alpha)=1$. In this setup, the tree-level $H^\pm W^\mp h$ coupling vanishes, while the decay $H^\pm\to W^\pm H$ is kinematically closed. The $H^\pm W^\mp A$ coupling, on the other hand, remains unsuppressed~\cite{Coleppa:2014cca,Sanyal:2019xcp,Mondal:2023wib,Sanyal:2023pfs}. 
We note that our assumption of the alignment limit of $s_{\beta-\alpha} = 1$ and mass degeneracy of $m_{H^\pm} \simeq m_H$ is justified in Fig.~\ref{fig:mA-tb}.

For $m_{H^\pm}>m_A+m_W$, the decay $H^\pm\to W^\pm A$ proceeds as a two-body process and dominates as the available phase space grows. Below the threshold, it proceeds through an off-shell $W$ boson, $H^\pm\to W^{\pm * }A\to Af\bar f'$, and is suppressed by the three-body phase space and the off-shell $W$ propagator. Hereafter, $W^{\pm(*)}A$ denotes both the on-shell and off-shell contributions. The competing fermionic modes, $H^\pm\to\tau^\pm\nu$, $H^\pm \to cs, cb$, and $H^\pm\to t^{(*)} b$, scale as $\cot^2\beta$ in the type-I 2HDM. Consequently, increasing $\tan\beta$ suppresses these fermionic widths while leaving the $H^\pm W^\mp A$ coupling unchanged, allowing the bosonic decay to dominate over a substantial region of the parameter space.

The resulting branching ratios are shown in Fig.~\ref{fig:BRHpWA}. For $m_A=50$ (left panel) and $70$~GeV (middle panel), the two-body $W^\pm A$ thresholds occur below $tb$ threshold. Consequently, the bosonic channel  becomes dominant for moderate values of $\tan\beta$. This behavior is somewhat delayed for $m_A=70$~GeV, since the larger pseudoscalar mass keeps the $W$ boson off shell over a wider range of $m_{H^\pm}$.

The structure of the contours changes around the $tb$ threshold. Once $H^\pm\to t b$ becomes kinematically allowed, the additional fermionic width competes with the $H^\pm\to W^\pm A$ channel, reducing its branching fraction at small $\tan\beta$, where the fermionic couplings are largest. This produces the characteristic kink in the contours near the $tb$ threshold. At larger $m_{H^\pm}$, the increasing phase space available to the bosonic channel allows it to become dominant again. As $\tan\beta$ increases, the fermionic widths are further suppressed by their $\cot^2\beta$ dependence. As a result, $\mathrm{BR}(H^\pm\to W^\pm A)\gtrsim90\%$ is obtained over most part of the parameter space for both $m_A=50$ and $70$~GeV, while values above $99\%$ are reached over broad regions at larger $\tan\beta$.

The right panel in Fig.~\ref{fig:BRHpWA} shows branching fraction contrours for $m_A=110$~GeV. In this case the two-body $W^\pm A$ threshold lies above the $tb$ threshold. At lower $m_{H^\pm}$, the $W^{\pm*}A$ channel is phase-space suppressed, although it can still dominate when the competing fermionic widths are sufficiently small. Once the $tb$ channel becomes kinematically allowed, its partial width increases while the $WA$ channel remains off-shell or close to its two-body threshold. This leads to the strong suppression of the bosonic branching fraction in the region $m_{H^\pm}\simeq175$--$200$~GeV. With a further increase in $m_{H^\pm}$, the on-shell $W^\pm A$ channel opens and its branching fraction increases again. Compared with the lighter pseudoscalar cases, larger values of $\tan\beta$ or $m_{H^\pm}$ are therefore required to obtain a sizable bosonic branching fraction.

\section{Collider Study}
\label{sect:analysis}
 
As discussed in the previous section, the presence of a light pseudoscalar allows the bosonic decay
of the $H^{\pm}$  to dominate over a substantial region of the type-I 2HDM parameter space and we utilize this decay mode to search for the charged Higgs boson at the High-Luminosity LHC (HL-LHC).
In this analysis, we focus on the QCD-induced charged Higgs production and consider the signal process
\begin{equation}
    pp\to b\bar b\,W^\mp H^\pm  \to b\bar b\,W^\mp W^{\pm(*)} A \to 4b\,jj\,\ell\nu,
    \label{eq:collider-signal}
\end{equation}
where $\ell=e,\mu$.
The production process contains the double-resonant, single-resonant, and non-resonant top-quark contributions,
together with their interference, as discussed in the previous section.

To suppress the large QCD multijet background, we require one of the two $W$ bosons to decay leptonically while the 
other decays hadronically. Since the pseudoscalar dominantly decays into $b\bar b$, the signal gives rise to a final state 
containing four bottom quarks, one charged lepton, light jets, and missing transverse momentum.

The analysis is performed for proton--proton collisions at $\sqrt{s}=\text{13}$~TeV with an integrated luminosity
$\mathcal{L}=\text{3000}$~$\mathrm{fb}^{-1}$. We implement the type-I 2HDM in \texttt{FeynRules}~\cite{Alloul:2013bka} 
and generate tree-level signal events using \texttt{MadGraph5\_aMC@NLO v3.4.2}~\cite{Alwall:2011uj,Alwall:2014hca}. For event generation, we use \texttt{NNPDF30\_lo\_as\_0118}\cite{NNPDF:2017mvq} parton distribution function and the renormalisation $\mu_R$ and factorization $\mu_F$ scales are set to $\frac{1}{3}\sum_i \sqrt{m_i^2 + p^2_{T,i}}$ where $i$ runs over all final state particles used in the matrix element calculation. The generated parton-level events are passed to \texttt{Pythia-8.3}~\cite{Sjostrand:2006za,Sjostrand:2014zea} 
for parton showering and hadronization. Detector effects are simulated using 
\texttt{Delphes-3.5.0}~\cite{deFavereau:2013fsa}. For $b$--tagging, we used the $p_T$-dependent efficiencies corresponding to the `DeepCSV Medium' working point based on the $\sqrt{s}=13$\,TeV data from the CMS collaboration \cite{CMS:2017wtu} and updated the default CMS detector card accordingly in Delphes.
Jet clustering is performed using \texttt{FastJet}~\cite{Cacciari:2011ma} as implemented within the 
\texttt{Delphes} framework using anti-$k_t$ algorithm with $\Delta R = 0.4$. 
We rescaled the leading-order signal cross section to match the NLO QCD predictions for the three charged Higgs mass regimes~\cite{Degrande:2016hyf,LHCHiggsCrossSectionWorkingGroup:2016ypw}.

The dominant background arises from $t\bar t+\mathrm{jets}$ production~\cite{ATLAS:2024rcu}, where additional
$b$-tagged jets can originate from heavy-flavor radiation or from the mistagging of light
and charm jets. We used the 5FS \texttt{SM5F$\_$NLO} model file available in the \texttt{FeynRules} model database and matched the $t\bar t$ background events up to two jets. The inclusive $t\bar t$ background is normalized to the NNLO production
cross section, 833.9 pb,  obtained with \texttt{Top++2.0}~\cite{Czakon:2011xx}. 
Processes such as $W+\mathrm{jets}$ can also contribute as background through mistagging of 
light jets into $b$-jets. However, the requirement of four $b$-tagged jets strongly suppresses
these backgrounds and we have not considered these processes. Other top-quark induced backgrounds are also subdominant~\cite{ATLAS:2024rcu} and thus we neglect them in this analysis. 

\begin{figure}[t]
    \centering
    \includegraphics[width=0.48\textwidth]{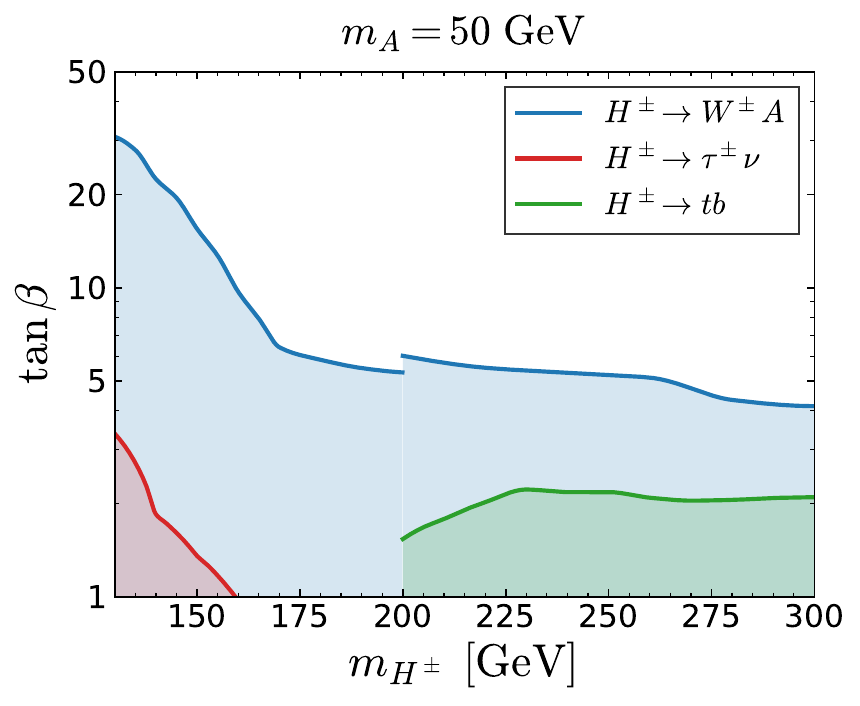}
    \hfill
    \includegraphics[width=0.48\textwidth]{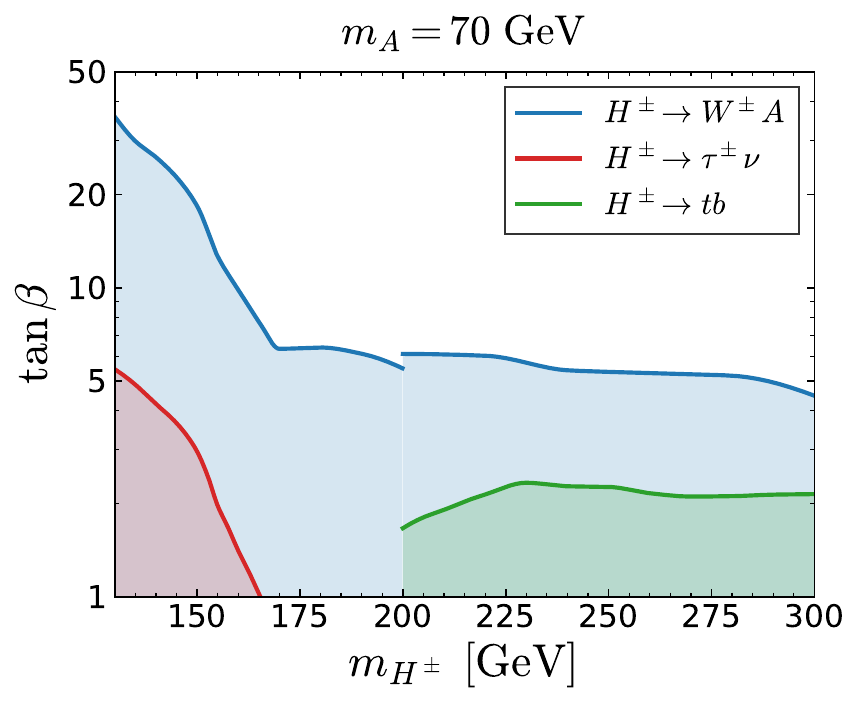}
    \\[1ex]
    \includegraphics[width=0.48\textwidth]{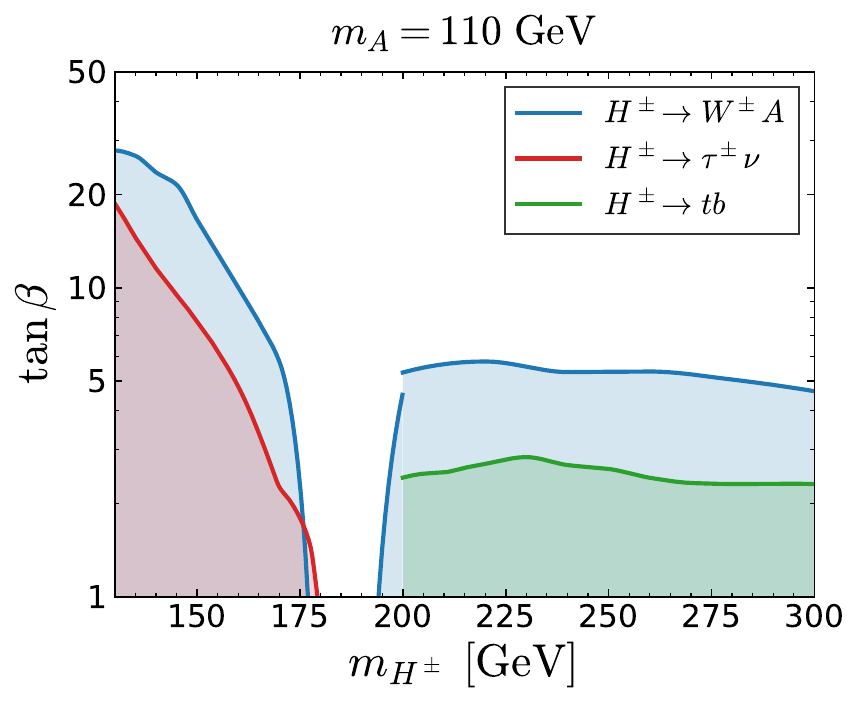}
    \caption{Projected HL-LHC sensitivity in the    $m_{H^\pm}$--$\tan\beta$ plane for $m_A=50$~GeV
    (upper left), $70$~GeV (upper right), and $110$~GeV (lower panel). The blue curves show the projected 
    $Z_{\rm stat}=2$ sensitivity for the $H^\pm\to W^{\pm(*)}A$ channel.  The red and green curves show the 
    projected LHC constraints from $H^\pm\to\tau^\pm\nu$~\cite{ATLAS:2024hya} and $H^\pm\to tb$~\cite{ATLAS:2021upq}, respectively.}
    \label{fig:result}
\end{figure}

For the signal and background event selection, we require at least four $b$-tagged jets and at least two light jets with
\begin{equation}
    p_T(b,j)>20~\mathrm{GeV}, \quad |\eta(b,j)|<2.5
\end{equation}
and one isolated electron or muon with
\begin{equation}
    p_T(\ell)>10~\mathrm{GeV}, \quad |\eta(\ell)|<2.5,
\end{equation}
along with minimal missing transverse energy, $E_T^{miss} > 10$ GeV.

We estimate the expected sensitivity using the Asimov significance~\cite{Cowan:2010js},
\begin{equation}
    Z_{\rm stat}
    =
    \sqrt{2\left[(s+b)\ln\left(1+\frac{s}{b}\right)-s\right]},
    \label{eq:significance_stat}
\end{equation}
where $s$ and $b$ denote the expected signal and background event yields after the selection criteria.
For each value of $m_{H^\pm}$, the dependence on $\tan\beta$ is incorporated through
the charged Higgs production cross section and the relevant decay branching fractions.
We define the projected $2\sigma$ sensitivity as the largest value of $\tan\beta$
for which $Z_{\rm stat}\geq2$.

The projected sensitivity of the $H^\pm \to W^\pm A$ channel is shown by the blue curves in Fig.~\ref{fig:result}.
For $m_A=50$ and $70$~GeV, the large charged Higgs production rate through top decay
provides sensitivity up to $\tan\beta\simeq30$--$35$ at the lower end of the charged Higgs mass range.
The reach decreases rapidly as $m_{H^\pm}$ approaches the top-decay threshold and around the transition to the heavy charged Higgs region, the sensitivity is reduced to
$\tan\beta\simeq5$--$6$ and remains around in the range
$\tan\beta\simeq4$--$6$ over most of the masses considered.
The small discontinuity around $m_{H^\pm}=200$~GeV originates from the change
from the 4FS to the 5FS discussed in the previous section.

The behavior is slightly different for $m_A=110$~GeV where the sensitivity falls sharply around
$m_{H^\pm}\simeq180$--$195$~GeV.  As discussed previously, for this pseudoscalar mass, the $tb$~channel 
opens before the on-shell $W^\pm A$ channel becomes accessible. The resulting increase of the fermionic width strongly suppresses $\mathrm{BR}(H^\pm\to W^{\pm(*)}A)$ and consequently the signal rate becomes low. Once the on-shell $W^\pm A$ decay becomes kinematically accessible, the sensitivity recovers, reaching $\tan\beta\simeq5$--$6$ above $m_{H^\pm}\simeq200$~GeV and decreasing gradually with increasing charged Higgs mass.

To illustrate the complementarity of the proposed search,
we also show in Fig.~\ref{fig:result} the HL-LHC projection of the existing LHC constraints from the fermionic channels
$H^\pm\to\tau^\pm\nu$~\cite{ATLAS:2024hya} and $H^\pm\to tb$~\cite{ATLAS:2021upq}.
To obtain the HL-LHC projected limits from the $\tau\nu$ and $tb$ channels, 
we employ the following naive luminosity scaling:
\begin{equation}
    (\sigma \times \mathrm{BR})^{3000\,\rm{fb}^{-1}}_{\rm{limit}} = (\sigma \times \rm{BR})^{140\,\rm{fb}^{-1}}_{\rm{limit}}\sqrt{\frac{140}{3000}}.
\end{equation}
For $m_A=50$ and $70$~GeV, these searches mainly constrain the low-$\tan\beta$ region,
while the $W^{\pm(*)}A$ channel remains sensitive to substantially larger values of $\tan\beta$.
In the heavy charged Higgs region, the $tb$ search typically probes values only around
$\tan\beta\sim2$, much below the reach of the bosonic channel.
For $m_A=110$~GeV, the $\tau\nu$ constraint becomes considerably stronger in the light charged Higgs region due to the suppression of the $W^{\pm(*)}A$ mode below its two-body threshold.
Once the bosonic channel becomes dominant, the conventional fermionic searches lose sensitivity rapidly with increasing $\tan\beta$.

The complementarity between the bosonic and fermionic searches is controlled by the charged Higgs decay pattern. 
 In the region where $H^\pm\to W^{\pm(*)}A$ dominates the total charged Higgs width, its branching fraction approaches unity, whereas the fermionic branching fractions inherit the $1/\tan^2\beta$ suppression of the corresponding partial widths. Consequently,
\begin{equation}
    \mathrm{BR}(H^\pm\to f f') \propto \frac{1}{\tan^2\beta},
    \qquad
    \mathrm{BR}(H^\pm\to W^{\pm(*)}A)\approx 1 .
\end{equation}
Since the QCD production cross section itself decreases approximately as $1/\tan^2\beta$, the fermionic signal rate is suppressed as $1/\tan^4\beta$, while the bosonic signal rate retains only the $1/\tan^2\beta$ suppression from production. The $H^\pm\to W^{\pm(*)}A$ channel therefore remains sensitive to larger values of $\tan\beta$, where the conventional fermionic decay modes are strongly suppressed.

\section{Conclusion}
 \label{sect:concl}
The discovery of a charged Higgs boson would provide unambiguous evidence for an extended Higgs sector beyond the SM. In particular, the type-I 2HDM offers an interesting framework in which a relatively light charged Higgs boson can remain compatible with existing flavor physics and collider constraints. While conventional searches have predominantly focused on fermionic charged Higgs decays, bosonic decay channels offer complementary and potentially robust approach for exploring the extended scalar sector.

In this work, we have investigated the prospects for probing the charged Higgs boson through its bosonic $H^\pm\to W^{\pm(*)} A$ decay at the LHC. We first performed a comprehensive scan of the type-I 2HDM parameter space, imposing theoretical constraints, electroweak precision observables, flavor constraints, direct search limits on  extra Higgs bosons and the measured properties of the observed 125 GeV Higgs boson. 
We subsequently studied  charged Higgs production and decay, highlighting the importance of the $H^\pm\to W^{\pm(*)}A$ channel. 
We have shown that a light pseudoscalar $A$ can substantially modify the charged Higgs phenomenology. In particular, $H^\pm\to W^{\pm(*)}A$ can become the dominant decay mode in regions of parameter space, namely for moderate-to-large $\tan\beta$.
We then performed a dedicated collider analysis of the charged Higgs signal with a  final state of $4b+2j +\ell\nu$,  
taking into account the relevant SM backgrounds and applying appropriate event-selection criteria to assess the sensitivity of the LHC to this signature.

Our study demonstrates that the bosonic $H^\pm\to W^{\pm(*)}A$ decay channel provides a promising  complementary probe of the type-I 2HDM, particularly in regions where conventional searches based on fermionic charged Higgs decays have limited sensitivity. 
A dedicated experimental search at the LHC for this signature could therefore provide an important opportunity to probe light pseudoscalar and charged Higgs bosons, opening a complementary window onto regions of the 2HDM type-I that are challenging  to explore through  conventional fermionic decay channels.

\section*{Acknowledgements}
AA gratefully acknowledges the hospitality of LAPTh, Annecy, where part of this work was carried
out during his sabbatical leave.
MK is supported by the National Science and Technology Council of Taiwan under grant No. 114-2639-M-002-006-ASP. TM is supported by Anusandhan National Research Foundation (ANRF), Government of India, under the Prime Minister Early Career Research Grant (Grant No. ANRF/ECRG/2025/002387/PMS). PS is supported by Basic Science Research Program through the National Research Foundation of Korea (NRF) funded by the Ministry
of Education through the Center for Quantum Spacetime (CQUeST) of Sogang University with
grant number RS-2020-NR049598 and by the Ministry of Science and ICT with grant number
RS-2025-24523022.

\bibliographystyle{JHEP}
\bibliography{references}

				
\end{document}